\documentclass[a4paper,twocolumn]{esapub2005} 
\usepackage{graphicx}

\usepackage{booktabs}

\usepackage[binary-units]{siunitx}
\usepackage[acronym]{glossaries}
\newacronym{lro}{LRO}{Lunar Reconnaissance Orbiter}
\newacronym{lroc}{LROC}{Lunar Reconnaissance Orbiter Camera}
\newacronym{pangu}{PANGU}{Planet and Asteroid Natural Scene Generation Utility}
\newacronym{nn}{NN}{neural networks}
\newacronym{svm}{SVM}{support vector machine}
\newacronym{cstm}{CSTM}{continuously scalable template model}
\newacronym{esa}{ESA}{European Space Agency}
\newacronym{nac}{NAC}{narrow-angle camera}
\newacronym{wac}{WAC}{wide-angle camera}
\newacronym{cnn}{CNN}{convolutional neural network}
\newacronym{gan}{GAN}{generative adverserial network}
\newacronym{ldm}{LDM}{latent diffusion model}

\usepackage{tikz}
\usetikzlibrary{shapes,arrows,positioning}

\usepackage{amsmath}
\usepackage{amsfonts}

\usepackage{hyperref}
\newcommand\rurl[1]{%
\texttt{\href{http://#1}{\nolinkurl{#1}}}
}

\usepackage{cleveref}
\crefname{table}{Tab.}{Tabs.}
\crefname{figure}{Fig.}{Figs.}
\crefname{section}{Sec.}{Secs.}
\crefname{equation}{Eq.}{Eqs.}

\usepackage{cuted}
\usepackage{caption}

\begin{document}

\title{\Large \bf LROC-PANGU-GAN: Closing the Simulation Gap in Learning Crater Segmentation with Planetary Simulators}
\author{
Jaewon La$^1$, Jaime Phadke$^1$,
Matt Hutton$^4$,
Marius Schwinning$^4$,
Gabriele De Canio$^4$,
Florian Renk$^4$,
Lars Kunze$^2$, and Matthew Gadd$^3$\\
$^1$Balliol College, $^2$Cognitive Robotics Group (CRG), $^3$Mobile Robotics Group (MRG), University of Oxford\\
\texttt{\{lars,mattgadd\}@robots.ox.ac.uk}\\
$^4$European Space Operations Centre, European Space Agency (ESA)\\
\texttt{Florian.Renk@esa.int}
\thanks{
This work was supported by the Human-Machine Collaboration Programme of the University of Oxford and Amazon Web Services (AWS).
Lars Kunze is supported by EPSRC Project RAILS (EP/W011344/1).
Matthew Gadd is supported by EPSRC Programme Grant ``From Sensing to Collaboration'' (EP/V000748/1).
}
}
\maketitle

\begin{strip}
\centering
\vspace{-2.5cm}
\includegraphics[width=\textwidth]{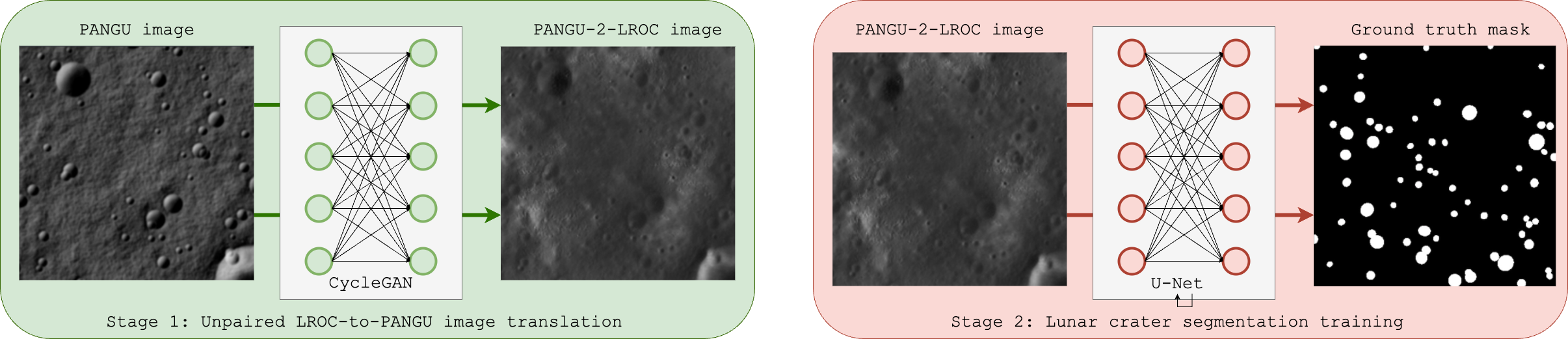}
\captionof{figure}{Simplified overview of our two-stage system.
\texttt{Stage 1} is the core contribution of this work, while \texttt{Stage 2} is used to demonstrate the usefulness of the synthesised outputs.
By transforming synthetic images of the lunar landscape from the \acrlong{pangu} simulator into images which, although synthesised, more faithfully capture the nuances and attributes of the moon's surface, we produce more realistic-looking lunar data.
This is important for training subsequent models e.g. for segmenting and detecting lunar craters, since our synthesised data is derived from precise \acrlong{pangu} crater locations, which real crater databases cannot guarantee.
}
\label{fig:lroc-pangu-gan}
\vspace{.5cm}
\end{strip}

\begin{abstract}
It is critical for probes landing on foreign planetary bodies to be able to robustly identify and avoid hazards -- as, for example, steep cliffs or deep craters can pose significant risks to a probe's landing and operational success.
Recent applications of deep learning to this problem show promising results.
These models are, however, often learned with explicit supervision over annotated datasets.
These human-labelled crater databases, such as from the \gls{lroc}, may lack in consistency and quality, undermining model performance -- as incomplete and/or inaccurate labels introduce noise into the supervisory signal, which encourages the model to learn incorrect associations and results in the model making unreliable predictions.
Physics-based simulators, such as the \acrlong{pangu}, have, in contrast, perfect ground truth, as the internal state that they use to render scenes is known with exactness.
However, they introduce a serious simulation-to-real domain gap -- because of fundamental differences between the simulated environment and the real-world arising from modelling assumptions, unaccounted-for physical interactions, environmental variability, etc.
Therefore, models trained on their outputs suffer when deployed in the face of realism they have not encountered in their training data distributions.
In this paper, we therefore introduce a system to close this ``realism'' gap while retaining label fidelity.
We train a CycleGAN model to synthesise \gls{lroc} from \gls{pangu} images.
We show that these improve the training of a downstream crater segmentation network, with segmentation performance on a test set of real \gls{lroc} images improved as compared to using only simulated \gls{pangu} images.
This will in the future allow researchers to more robustly test in-the-loop autonomous systems when tested in simulators such as \gls{pangu}.
\end{abstract}

\begin{figure*}[!h]
\centering
\includegraphics[width=0.85\textwidth]{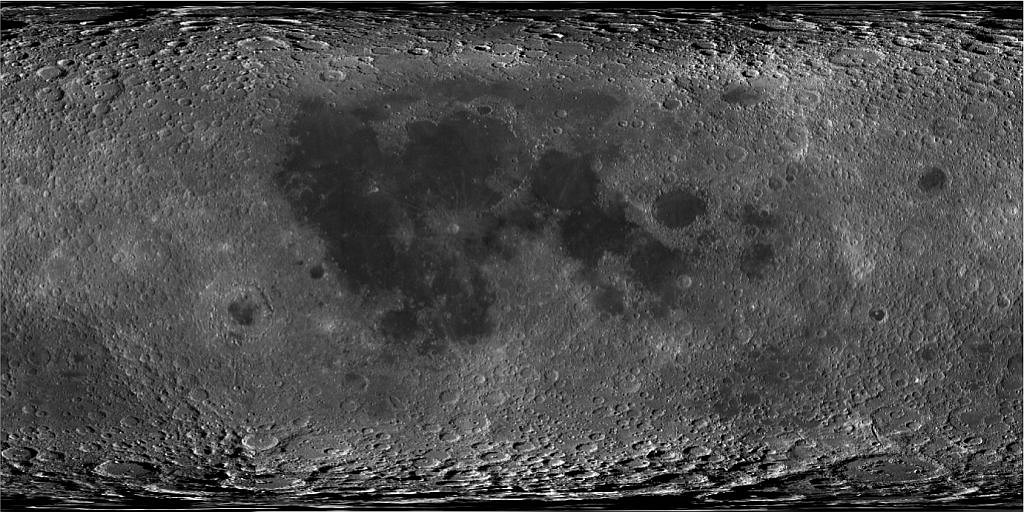}
\caption{
The \acrfull{lroc} dataset used in this investigation was obtained through the NASA and \gls{esa}
databases.
The dataset consists of a very high resolution image of the entire lunar
surface, which was generated by stitching together thousands of images taken by the \gls{lro}, with corresponding document containing the location, size, and
the approximate shape of the craters. The approximate resolution of the \gls{lroc} image is
around \SI{100}{\metre} per pixel.
In the associated \textit{Lunar Crater Database}~\cite{robbins2016developing}, labels of craters were generated by
human inspection, i.e. human volunteers hand-labelled each crater in the image.
}
\label{fig:LROC_Image}
\end{figure*}

\section{Introduction}

Accurate representation of extraterrestrial landscapes is important for landing~\cite{maass2020crater} and navigation~\cite{downes2020lunar} -- for example, terrain understanding can improve the precision of a spacecraft navigation by positioning relative to large-scale features or may inform terrestrial missions~\cite{balme20192016} after landing -- as well as scientific understanding of the geology and evolution of planets and moons~\cite{lagain2021model}.

In this area, data from the the \acrfull{lroc}~\cite{robinson2010lunar}, a high-resolution imaging system on board the \acrfull{lro} spacecraft~\cite{vondrak2010lunar}, is increasingly being used to learn deep models for crater detection~\cite{kerner2019toward,seo2020deep,del2022deep,seo2021lunar,chen2022cnn,la2023yololens,downes2020deep,giannakis2023deep,seo2021lunar,jin2019small}.
In these, \glspl{cnn} are well-suited to learning robustness to variability in cameras arises from illumination conditions, sensor sensitivity, noise, camera settings, and lens characteristics.
The lunar surface as imaged by this sensor is shown in~\cref{fig:LROC_Image}.

\begin{figure*}[!h]
\centering
\includegraphics[width=0.7\textwidth]{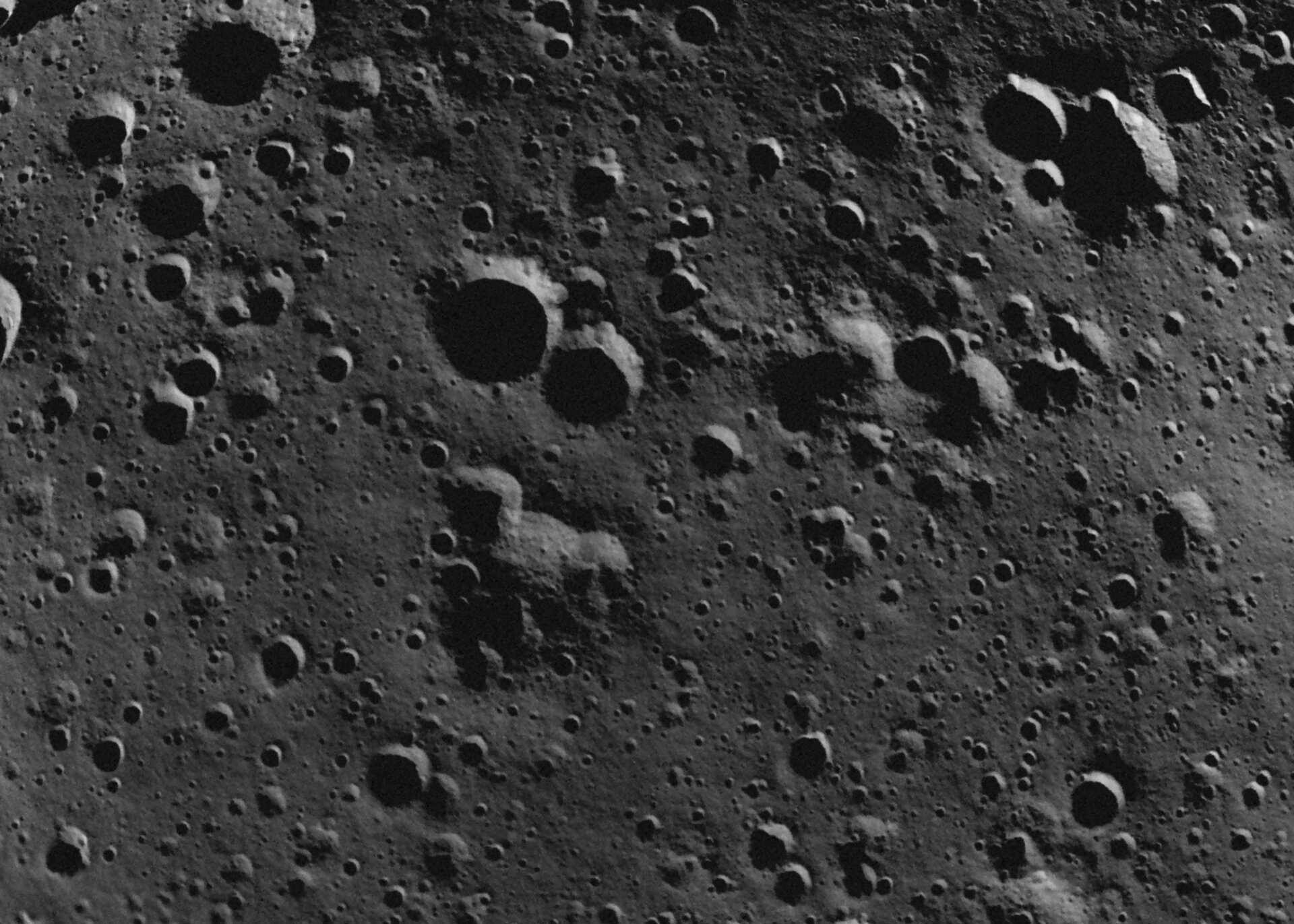}
\caption{
Sample \gls{pangu} image, taken from 
\rurl{www.esa.int/Enabling_Support/Space_Engineering_Technology/Shaping_the_Future/Planet_and_Asteroid_Natural_Scene_Generation_Utility_PANGU_Tool_Enhancement}.
We chose \gls{pangu} as our baseline
simulator due to its ease of use and the numerous options it has to generate various terrains
and its widespread adoption in terrain simulation.
As the default parameters are set
to resemble the lunar surface, parameters for the distribution in size and age of the craters
were left untouched.
One issue with the \gls{pangu}
outputs is the artificially well-defined crater edges.
This can be somewhat remedied by adjusting the crater edge decay option in the \gls{pangu} interface.
However, manual fine-tuning of this kind can be avoided in a learned approach, which can additionally capture other non-intuitive factors which are not modelled by the procedural simulator.
}
\label{fig:PANGU2_Malapert_crater_pillars}
\end{figure*}

Alongside this, the \acrlong{pangu}~\cite{parkes2004planet} and its extensions~\cite{mccrum2010mars} are popular for in-the-loop testing of spacecraft systems in simulation as well as as a source of data for machine learned systems~\cite{dubois2009testing,willis2016reinforcement,woicke2016stereo,martin2018testing,sikorski2021event,paschero2022assessment,chekakta2022robust,zhou2023crater,izzo2011constant}.
Here, simulators are valuable for training deep models as they provide abundant data, are controllable and cost-effective, and allow rapid prototyping -- i.e. they are safe and efficient.
The lunar surface as simulated by \gls{pangu} is shown in~\cref{fig:PANGU2_Malapert_crater_pillars}, which -- while very detailed -- is readily identifiable as simulated. 

Comparing~\cref{fig:LROC_Image,fig:PANGU2_Malapert_crater_pillars}, however, there remains a gap in realism.
For this, models trained with simulated data which is synthetically altered to be more realistic perform better than models trained with data generated by simulators as well as with real data that is altered with classical image-processing techniques (blurring, flipping, etc)~\cite{blumenkamp2022closing}.
Here, learned augmentations can achieve more complex appearance transformations that classical augmentations do not capture -- useful when data features intricate patterns or features.

\begin{figure*}
\centering
\includegraphics[width=0.75\textwidth]{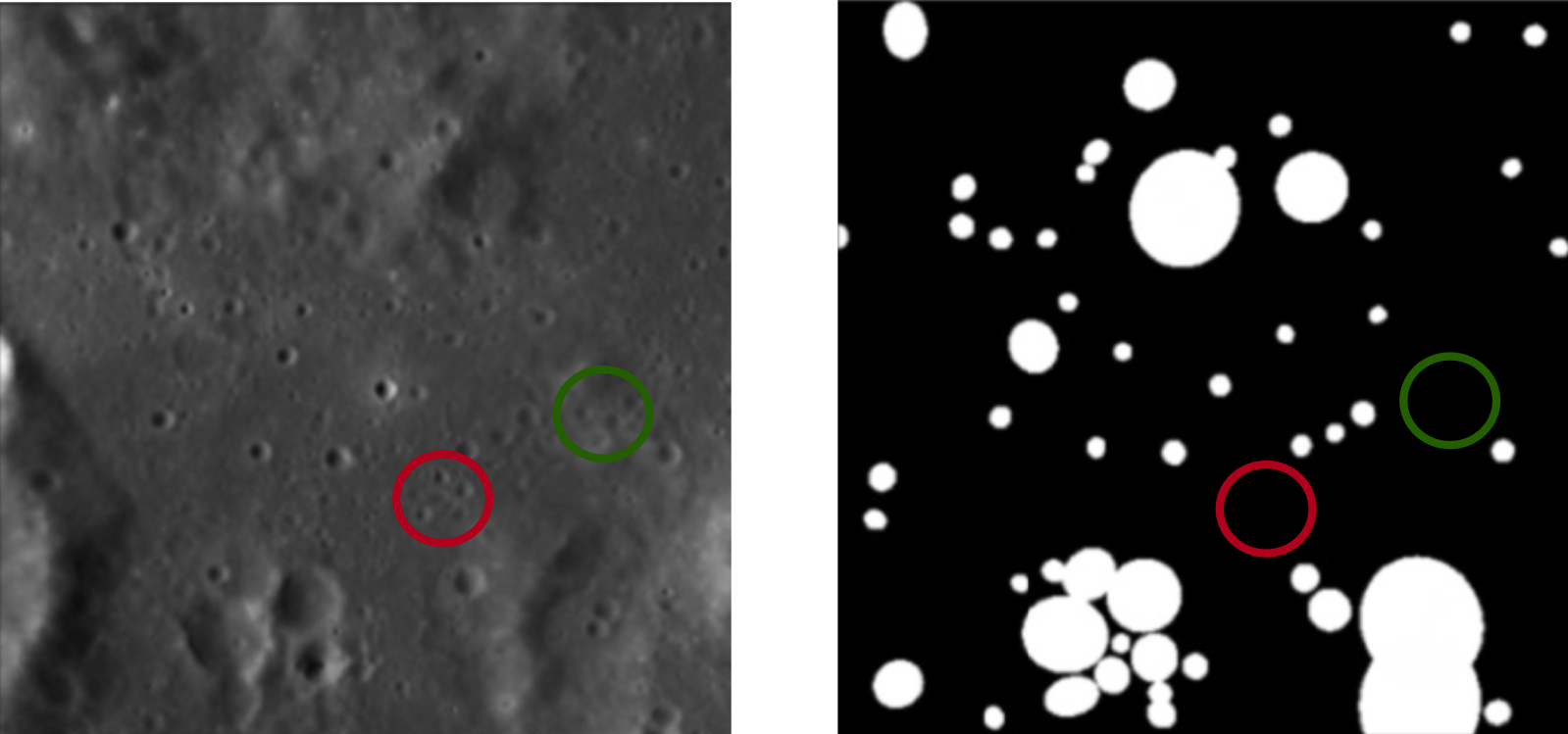}
\caption{
\textit{Lunar Crater Database} sample (left) and corresponding binary segmentation mask (right).
Some small craters are missing (e.g. in the top right of the frame) -- we indicate two such clusters (green and red), but on inspection find many other missing craters of a similar size.
This is determintal when optimising \glspl{nn} or other supervised machine learning models. 
}
\label{fig:MissingCraters}
\end{figure*}

Therefore, in this work, we have developed an approach to generating realistic lunar surface images using a CycleGAN~\cite{zhu2017unpaired}.
By converting ``fake'' images from the \gls{pangu} simulator into highly realistic images that accurately represent the features and characteristics of the lunar surface, we are capable of generating more realistic data.
This is essential for training downstream segmentation and detection models, as our generated data is accompanied by perfect \gls{pangu} ground truth, which are laborious and expensive to perform by hand and are imperfect or incomplete for the \textit{Lunar Crater Database}~\cite{robbins2016developing}\footnote{\rurl{astrogeology.usgs.gov/search/map/Moon/Research/Craters/lunar_crater_database_robbins_2018}} annotation of \gls{lroc} dataset, as motivated in~\cref{fig:MissingCraters}.
Here, imperfection in human labels is often due to \textit{subjectivity} (different ways of interpreting visual information), ambiguity (images with complex or ambiguous content), etc.

\section{Related Work}

Here we discuss existing crater identification systems to identify works for which our method for synthesising more realistic lunar surface imagery will be beneficial -- in training and testing.

\textbf{Manual identification} involves classification of lunar craters based on their size and morphology~\cite{robbins2016developing,povilaitis2018crater}, also performed in~\cite{lagain2021mars} for Mars impact craters by a larger team.
However, this method is \textit{time-consuming} can be \textit{subjective} (prone to human error), and does not scale well to large projects.

\textbf{Digital image processing} with classical machine vision techniques reduces this effort, which may involve edge detection and ellipse fitting~\cite{kim2005automated}.
These methods must often be \textit{tuned by hand} and are especially \textit{prone to image noise}.

\textbf{Machine learning} techniques, instead, are optimised over examples to distinguish data based characteristic features.
Early work in this area applied \glspl{svm}~\cite{ding2013novel} and \glspl{cstm}~\cite{wetzler2005learning}.
These methods, however, still require \textit{feature engineering}.

\textbf{\Acrlong{nn}}, in contrast, learn hierarchical representations directly from large amounts of data.
This has been applied to crater detection in various ways, including for bounding box prediction~\cite{jin2019small,seo2021lunar,la2023yololens}, crater rim segmentation~\cite{silburt2019lunar,downes2020deep}, and segmentation of the entire crater~\cite{del2022deep,giannakis2023deep}.

In applying these techniques to lunar crater detection, the issue of \textbf{incomplete labelled sets} over a range of lunar crater databases is discussed in~\cite{zang2021semi}.
Therefore, in this work, we turn to simulators for perfect label sets, and propose learning to apply realism to the simulated samples from unpaired domain adapation.

\textbf{Closest to our work} is~\cite{proencca2020deep,perez2023spacecraft}.
In~\cite{proencca2020deep}, sim-to-real transfer involves changing image exposure and contrast, additive white Gaussian noise, and blurring -- whereas our sim-to-real transfer is learned onto the real data distribution.
In~\cite{perez2023spacecraft}, domain adaptation is performed at the feature level, meaning that there is no sim-to-real output at the sensor level as in our work. 
Both~\cite{proencca2020deep,perez2023spacecraft} focus on spacecraft pose estimation -- with views of orbiting craft -- rather than bird's eye views of planetary landscapes.

Finally, while this initial proof-of-concept investigation was carried out with \glspl{gan}, future work will leverage improvements from recent \glspl{ldm}~\cite{rombach2022high} which offer better control over the generation process and exhibit improved convergence properties. 

\section{Method}

Our system is as illustrated in~\cref{fig:lroc-pangu-gan}.
It is a two-stage process, consisting of first training an image translation model and then using the outputs of the image translation model to train a crater segmentation model.
Therefore we are demonstrating the utility of our synthesised images in a \textit{downstream task}.
Other downstream tasks which could benefit from this more realistic source of training data may include learned pose estimation for autonomous space landers~\cite{chekakta2022robust}.

\subsection{PANGU}
\label{sec:pangu}

The \acrfull{pangu}~\cite{parkes2004planet} is a simulator used to generate synthetic images of planetary surfaces and asteroids
It is often used in space mission planning and analysis, and it is developed and maintained by \gls{esa}.
\gls{pangu} uses a combination of high-resolution topographic data, multispectral imaging data, and other parameters to create 3D models of planetary surfaces and asteroids, and then generates images of the landscapes from various viewpoints.
The resulting images can be used to simulate the appearance of a spacecraft's view during a flyby, landing, or other mission phases.

One issue with the \gls{pangu} dataset is the artificially well-defined crater edges.
While this can be somewhat remedied by tuning the crater edge decay option on \gls{pangu}, it still does not solve the issues of the differing terrain appearances, and also suggests that there are other non-intuitive factors at play. Therefore, it is more beneficial to take a learned approach to match the general appearance of the terrain to a realistic corpus such as \gls{lroc}.

In our work, a dataset was generated using the \gls{pangu} software.
Large tiles of size $8192$ by $8192$ pixels were made and sliced into smaller tiles of size $416$ by $416$ pixels.
With an overlap of $208$ pixels between subsequent tiles, each large tile yielded a total of $1369$ smaller tiles.
The maximum crater side-length was limited to $200$ pixels to ensure that no craters are bigger than the tile itself.

\begin{figure*}
\centering
\includegraphics[width=0.35\textwidth]{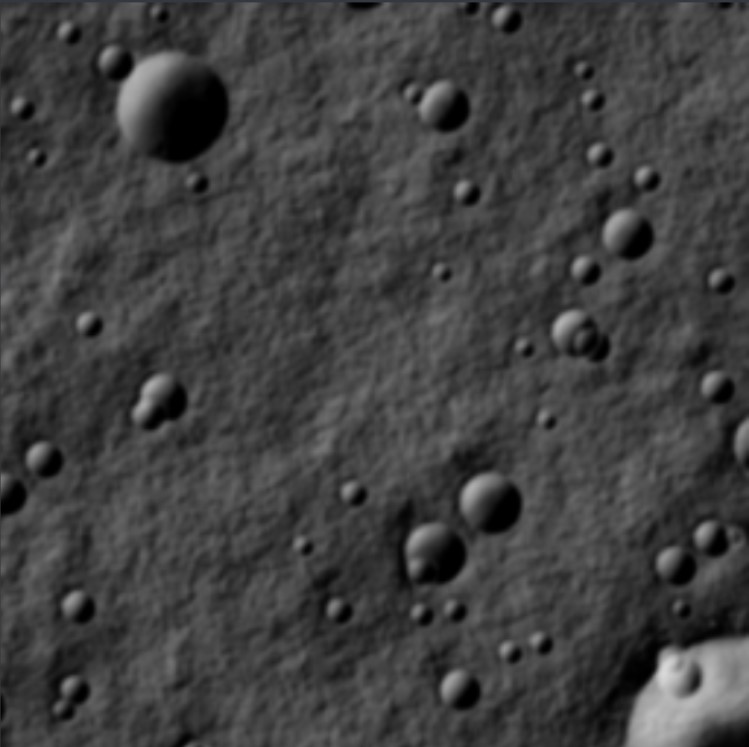}
\includegraphics[width=0.35\textwidth]{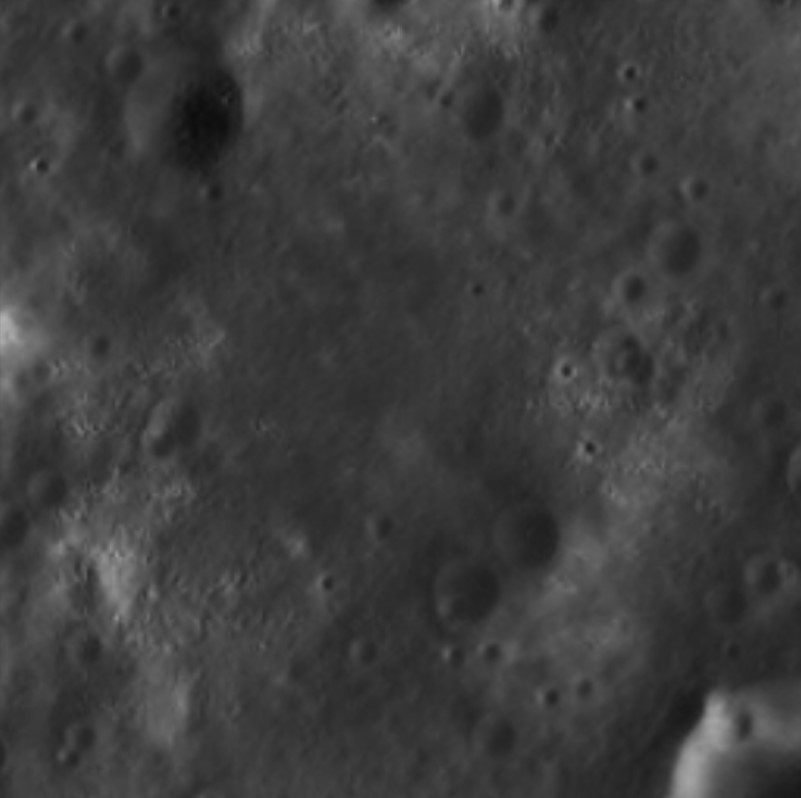}
\caption{
The left side of the image shows a fake lunar surface image that is clearly computer-generated and lacks the realism and detail of a genuine image. On the right side of the image, the synthesized lunar image created using the CycleGAN model is highly realistic and visually striking. The model has successfully captured the textures, shadows, and other characteristics of the Moon's surface
}
\label{fig:Original_pangu_Transferred_Pangu}
\end{figure*}

\subsection{LROC}
\label{sec:lroc}

The \acrfull{lroc} is a high-resolution imaging system on board the \acrlong{lro} spacecraft.
Launched by NASA in $2009$, \gls{lro} is a robotic mission designed to map and study the surface of the Moon.

The \gls{lroc} system consists of three separate cameras: two \glspl{nac} and one \gls{wac}.
The \glspl{nac} capture images at a resolution of \SI{0.5}{\metre} per pixel, while the \gls{wac} captures images at a resolution of \SI{100}{\metre} per pixel.

In our work, as broad views suffice to capture the general appearance of the lunar surface and considering it is this visual theme we want to synthesise, we use the \gls{wac}.
To conform with the resolution in~\cref{sec:pangu}, images from this camera were sliced into tiles of size $416\times416$ pixels.
These cover the full range of longitudes on the moon and the range of latitudes between $\pm\SI{60}{\degree}$.
Each tile covers an area of approximately \SI{1730}{\kilo\metre\squared}.

\subsection{LROC2PANGU (CycleGAN)}
\label{sec:cyclegan}

CycleGAN~\cite{zhu2017unpaired} enables unsupervised image-to-image translation between two domains -- it learns to capture style and content information from the input images.
It does this without the need for paired training data.
Specifically, CycleGAN enforces cycle consistency, which ensures that when an image is translated from one domain to another and then back, it should remain similar to the original image.

Our CycleGAN\footnote{
With implementation taken from \rurl{github.com/junyanz/pytorch-CycleGAN-and-pix2pix}
} uses generator networks (which perform the image translation -- from one style to another) formed by $9$ contracting and $9$ expanding ResNet~\cite{he2016deep} blocks, each block of which in brief consists of either a convolutional (contracting) or deconvolutional (expanding) layer, and which for ResNet are connected via shortcuts to other layers in order to avoid vanishingly small gradients deep in the network.

The discriminators (which the generators must fool into not being able to tell synthesised styles from original images) are formed by $3$ convolutional layers with non-linear ReLU activations. 

\subsection{Crater segmentation (U-Net)}
\label{sec:unet}

Given the success shown in Jackson \textit{et al}~\cite{silburt2019lunar} for crater detection, a U-Net model was chosen, although our core contribution is in improving the realism of lunar simulators and our method is applicable for other segmentation models. 

U-Net features a contracting path to capture context and a symmetric expanding path to precisely localise object boundaries, making it highly effective for pixel-level image segmentation.
Specifically, the output of the expanding path is supervised by (optimised to produce) binary segmentation masks as in~\cref{fig:MissingCraters,fig:lroc-pangu-gan,fig:qual}, with aerial views input to the contracting path.

Our U-Net implementation\footnote{
Similar to \rurl{github.com/milesial/Pytorch-UNet}
} consists of $4$ down-sampling blocks, each using an instance-normalised convolutional layer.
These blocks progressively reduce the input spatial resolution and produce deeper feature representations.
This is followed by $4$ up-sampling blocks that progressively up-sample the spatial resolution and return to the original image channel depth.

Our implementation of the U-Net model was trained over $30$ epochs with a batch size of $15$ and learning rate of $1\mathrm{e}^{-4}$.
The model is mainly comprised of a series of $3\times3$ convolutions, with a stride of $1$ and $1$ pixel of padding.
We use $5$ convolutional layers for both the contraction and expansion path.

\section{Experimental Setup}
\label{sec:exp}

\textbf{Datasets} Our training and testing datasets contain craters smaller than \SI{16}{\kilo\metre} in radius so as to prevent cases where the entire tile is occupied by a single crater.
We consider it important to focus on these smaller craters as large craters would otherwise ``mask out'' many small craters.
All of the models were trained with approximately $10000$ images in the training dataset and $1000$ images in the validation dataset.

\textbf{Performance metrics} We use a range of metrics to determine the performance of the model.
\begin{enumerate}
\item \textit{Accuracy} is calculated as the proportion of the pixels that were predicted correctly in the entire image.
\item Intersection over Union \textit{(IoU)} is calculated by dividing the area of intersection between the predicted and ground truth regions by the area of their union.
\item \textit{Precision} indicates how many of the pixels predicted to be a crater were actually crater.
\item \textit{Recall} indicates what proportion of actual craters were predicted as crater.
\item $F_1$ is the harmonic mean of precision and recall.
\item \textit{Specificity}, or the true negative rate, quantifies how well a classifier can avoid false crater detections by correctly identifying the regions which are not craters.
\end{enumerate}

\section{Results}

In the following we compare segmentation models trained with only simulated examples against models trained with our sim-to-real transfer applied, as a proof that the resulting images are more useful for understanding the lunar surface.

\cref{fig:Original_pangu_Transferred_Pangu} shows (left) a \gls{pangu} sample which is then processed by our trained image translation network to appear more realistic (right).
Upon visual inspection we observe that craters themselves are preserved, but appear more realistically ``textured'' in the synthesised result.
Note that these are $416\times416$ samples as prepared in~\cref{sec:cyclegan}, not full-size tiles as in~\cref{fig:LROC_Image}.

\begin{figure*}
\centering
\includegraphics[width=0.318\textwidth]{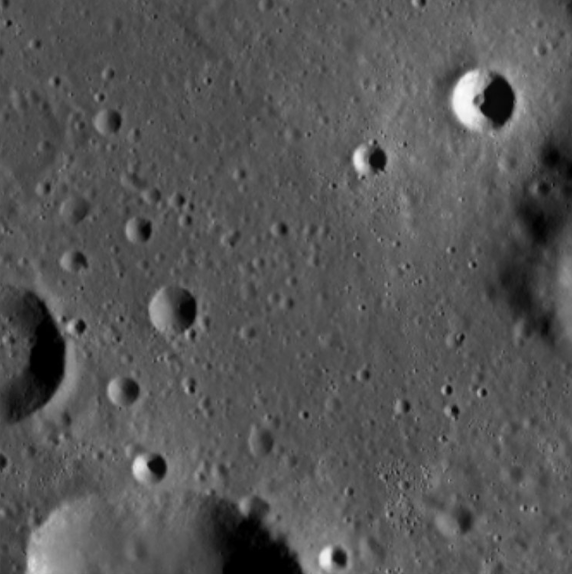}
\includegraphics[width=0.32\textwidth]{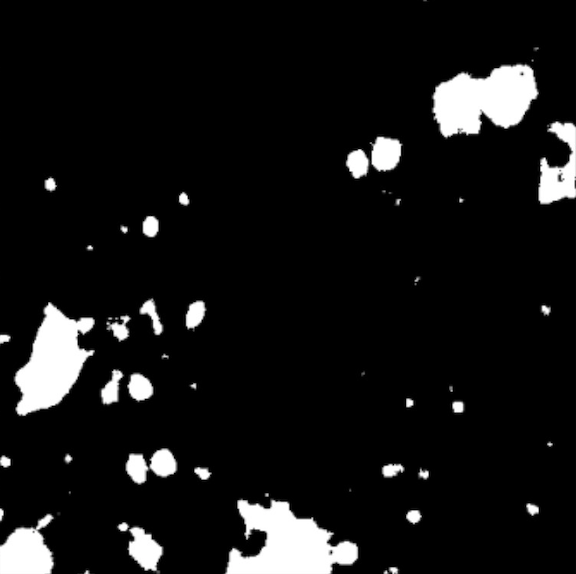}
\includegraphics[width=0.32\textwidth]{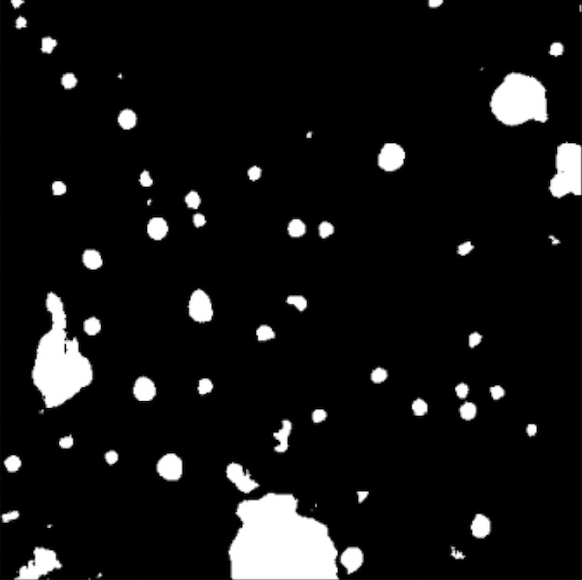}
\caption{\textit{Top}: Example image of PANGU trained model on LROC small crater dataset.
\textit{Bottom}: Example image of transferred PANGU trained model on LROC small crater dataset.}
\label{fig:qual}
\end{figure*}

\cref{fig:qual} then shows (middle) the crater segmentation mask predicted from an input \gls{lroc} image (left) when the U-Net model (\cref{sec:unet}) is only trained with \gls{pangu} images, as opposed to the predicted segmentation when U-Net is trained with our sim-to-real transfer applied to all training examples (right).
As can be seen, the model is better at identifying craters, with large craters more closely following the true outline (e.g. the top right round crater in particular -- with \gls{pangu}-only models predicting falsely a pair of craters) and, crucially, a much more comprehensive capture of the small craters.

\begin{table}[!h]
\centering
\caption{Crater detection performance metrics when training on simulated (PANGU) or synthesised (PANGU2LROC) data.\label{tab:metrics}}
\resizebox{\columnwidth}{!}{
\begin{tabular}{@{}l|cc@{}}
\centering
\textbf{Method} $\rightarrow$ & PANGU & PANGU2LROC (\textit{Ours})\\
\midrule
Accuracy [\SI{}{\percent}] $\uparrow$ & $89.59$ & \underline{$\mathbf{92.53}$} \\
$F_1$ [\SI{}{\percent}] $\uparrow$ & $15.29$ & \underline{$\mathbf{23.31}$}\\
IoU [\SI{}{\percent}] $\uparrow$ & $8.46$ & \underline{$\mathbf{13.81}$} \\
Precision [\SI{}{\percent}] $\uparrow$ & $12.71$ & \underline{$\mathbf{22.09}$} \\ 
Recall [\SI{}{\percent}] $\uparrow$ & $28.41$ & \underline{$\mathbf{33.13}$} \\
Specificity [\SI{}{\percent}] $\uparrow$ & $92.47$ & \underline{$\mathbf{95.37}$} \\
\end{tabular}
}
\end{table}

\cref{tab:metrics} lists quantitative segmentation performance metrics to support the qualitative proof of the advantage of our method in~\cref{fig:qual}.
These are averaged over the $1000$-image validation set, see~\cref{sec:exp}.
We observe improvements in all performance metrics, indicating that our sim-to-real transfer produces images more useful for training crater detection models than simulation alone.

\section{Conclusion}

We have presented a system for closing the realism gap for rendered images from planetary simulators.
We have proved the fidelity of its outputs by improving the training of a downstream lunar crater segmentation model.
This system will in the future be useful for more robust in-simulator testing of lunar operations, presenting autonomous and otherwise tasks with more realistic lunar data and scenarios.

\bibliographystyle{ieeetr}
\bibliography{biblio}

\end{document}